\documentclass[12pt]{article}

\def\half{\textstyle{\frac{1}{2}}}

\def\H{{\cal H}}

\def\p{\varphi}

\def\H{{\cal H}}

\def\l{\lambda}

\def\t{\textstyle}

\def\ra{\rightarrow}
\def\tint{{\textstyle\int}}

\def\hp{{\hat\pi}}
\def\hph{{\hat\varphi}}

\def\d{\partial}

\def\b{\begin{eqnarray*}}  
\def\e{\end{eqnarray*}}    
\def\bn{\begin{eqnarray}}  
\def\en{\end{eqnarray}}   

\def\<{\langle}
\def\>{\rangle}

\def\bk{\mathbf k}

\def\de{\delta}
\def\no{\nonumber}

\def\ds{d^s\!x}
\def\k{\kappa}

\def\v{\vskip1em}
\def\hk{\hat{\kappa}}

\def\{{\lbrace}
\def\hv{\hat{\varphi}}
\def\}{\rbrace}

\begin{document}

\title{Evidence for Expanding \\Quantum Field Theory}  
  \author{John R. Klauder\footnote{klauder@ufl.edu} \\
Department of Physics and Department of Mathematics  \\ 
University of Florida,   
Gainesville, FL 32611-8440}
\date{ }
\bibliographystyle{unsrt}

\maketitle 

\begin{abstract}
Present day quantum field theory (QFT) is founded on canonical quantization, which has served 
quite well, but also has led to several issues. The free field describing a free particle
(with no interaction term) can suddenly become nonrenormalizable the instant a suitable 
interaction term appears. For example, using canonical quantization, $\p^4_4$, has been 
deemed  a ``free" theory with no difference from a truly free field \cite{a,b,c}. Using the same model, affine  quantization has led to a truly interacting theory \cite{fk}.
This fact alone asserts that canonical {\it and } affine tools of quantization deserve to be open to their procedures together as a significant enlargement of QFT.
\end{abstract}
\section{Introduction}
\subsection{Classical starting point}
There are different ways to promote classical to quantum expressions that are useful.
For the classical, canonical Hamiltonian model, we have
  \bn   H(\pi,\p)= \tint \{ \half[ \pi(x)^2 + (\overrightarrow{\nabla}\p(x))^2+m^2 \p(x)^2] +g\,\p(x)^r\,\}\; d^s\!x 
  \label{67} \;, \en
  where $r\geq 2$ and $s\geq 3$.
  The ingredients in this expression are the classical field $\p(x)$ and the momentum field
  $\pi(x)$. These fields obey the Poisson bracket $ \{\p(x),\pi(y)\}=\delta^s(x-y)$.
  
  However, we can describe the same Hamiltonian in a different way. Let us choose the affine field $k(x)\equiv 
  \pi(x)\,\p(x)$, instead of the momentum field, but still keep $\p(x)$. However, it is necessary to keep $\p(x)\neq 0$ for otherwise if $\p(x)=0$, then $\pi(x)$ means nothing. Let us exam (\ref{67}),
  the same classical Hamiltonian, but now in the new coordinates, leading to
  \bn H'(\k,\p)=\tint \{\half[ \k(x)^2\,\p(x)^{-2}+(\overrightarrow{\nabla}\p(x))^2+m^2\,\p(x)^2] 
  +g\,\p(x)^r\}\; d^s\!x\;. \en
 This new set of fields leads to the Poisson bracket  $\{\p(x), \k(y)\} =\delta^s(x-y)\,\p(x)$.

\subsection{Quantum starting point}
Promotion of the fields $\p(x)\ra \hv(x)$ and $\pi(x)\ra \hp(x)$, leads to the traditional quantum expression for our Hamiltonian, which is given by
   \bn \H(\hp,\hv)=\tint \{ \half[\hp(x)^2 + (\overrightarrow{\nabla}\hv(x))^2 +m^2\hv(x)^2]+ g\, \hv(x)^r\,\}\; 
   d^s\!x\;.\en

Now, knowing that the classical variables were no longer the canonical choice but rather the
affine coordinates, and after the promotion of affine field variables to new quantum field variables, the new quantum Hamiltonian becomes
  \bn \H'(\hk,\hv)=\tint \{\half[ \hk(x) \, \hv(x)^{-2}\,\hk(x) +(\overrightarrow{\nabla}\hv(x))^2+ m^2 \hv(x)^2]
  +g\,\hv(x)^r\,\}\;
  d^s\!x \;. \en
  Do not worry about $\hv(x)^{-2}$ because we have already insisted  that $\p(x)\neq0$; hence $ \hv(x)\neq 0$. In a previous usage, which proved itself by modifying  $\hv(x)^{-2}\ra [\,\hv(x)^2+ \epsilon\,]^{-1}$, while $\epsilon =10^{-10}$ served as a safeguard \cite{22}.
  
  It is noteworthy that $\hk(x) \,\hv(x)^{-1/2}=0$, which, in Schr\"odinger's representation, leads to
  $\hk(x)= -i\hbar[\p(x)(\d/\d \p(x))+(\d/\d \p(x))\p(x)]/2$ and $\hv (x) =\p(x)$. Following  a suitable regularization process \cite{777}, this yields the stated result.
  
\subsection{Advantages of an affine quantization}\v
    Using the results of the previous sections we propose that $\hk(x) \, \Pi_y \p(y)^{-1/2}=0$
    which exposes our choice for  general wave functions as given 
    by $\Psi(\p)= \tint W(\p(x))\,\Pi_y \,\p(y)^{-1/2}\,d\p(x)$. A regularized version, using $x\ra \bk$ where 
    $\bk = a {\bf n} = a \times \{.., -2,-1,0,1,..\}^s $ of this expression looks like
    $\Psi(\p)= w(\p) \Pi_\bk (ba^s)^{1/2}(\p_\bk)^{-(1- 2ba^s)/2}$, where $ba^s$ is dimensionless and  $b\sim 1$. 
    
    We now take a Fourier transformation of the absolute square of our regulated wave
    function that looks like
 \bn F(f)=\Pi_\bk \,\tint \{ e^{if_\bk \p_\bk} |w(\p_\bk)|^2 (ba^s) \,|\p_\bk|^{-(1-2ba^s)}\,
 d\p_\bk\}\;.\label{6} \en
       Normalization ensures that if all $f_\bk =0$, then $F(0)=1$, which leads to
       \bn F(f)=\Pi_\bk \tint \{ 1- \tint(1- e^{if_\bk \p_\bk}) |w(\p_\bk)|^2 (ba^s)\,d\p_\bk/
       |\p_\bk|^{(1- 2ba^s)}\,\}\;.  \en
       Now, at last, we can let $a\ra 0$ to fix the Fourier transformation\footnote{Any change of $w(\p)$ due to $a\ra0$ is left implicit.}
  \bn F(f)= \exp\{ -b\tint \ds(1- e^{if(x)\,\p(x)}) |w(\p(x)|^2 d\p(x)/|\p(x)|\,\} \;. \label{2} \en
       
       Observe that the affine quantization his led to a Poisson distribution, which is the only other 
       term, besides a Gaussian expression, as dictated by the The Central Limit Theorem \cite{87}.
       Nevertheless, the same expression as in (\ref{2}) could have arisen when $g=0$, or 
       even when  $g\nearrow \,\searrow 0$, asserting that our final result is definitely not a Gaussian! Of significants is the fact that if the coupling $g$, or even the mass $m$, are smoothly changed, there are only continuous changes within $w (\p)$. Also, the fact that 
       $\hk(x)\,\p(x)^{-1/2}=0$, which is a dramatic change from canonical 
       theory's equivalent relation, i.e.,   $\hp(x)\, 1\!\!1=0$, makes a big difference; indeed, the factor $|\p_\bk|^{-(1-2ba^s)}$ in (\ref{6}) is the key to avoiding a Gaussian result. Aparently, this behavior of affine quantization adopts the least final domain 
       at the outset, which overcomes any threat of nonrenormalizability.\footnote{For those who wish  
       to learn more about affine quantization see \cite{777}. For beginners, canonical quantization 
       deals with the harmonic oscillator, but the half-harmonic oscillator requires affine 
       quantization \cite{32}.}
      
      \section{Summary}
      We have obtained a continuous, fully regularized, expression that implicitly involves a large 
      sample of quantum field models. The application of affine quantization, but not canonical 
      quantization, has offered us a treasure of interest that presently rests in the Fourier representation space. To understand the 
      physics needed to clarify our results requires a second Fourier transformation back into the 
      original space of the classical field, here given by $\p(x)$. That issue is purely a 
      mathematical task, and the implications of such an effort are certainly of great interest!

\end{document}